\documentclass[12pt,a4paper]{article}
\usepackage[pdfstartview=FitH,colorlinks=true,linkcolor=black,anchorcolor=black,citecolor=black,urlcolor=black]{hyperref}
\usepackage[english]{babel}
\usepackage{amsmath,amssymb,titling,authblk}
\usepackage{slashed}
\usepackage{amsmath}
\usepackage{amscd}
\usepackage{mathrsfs}
\usepackage[normalem]{ulem}
\usepackage{appendix}
\usepackage{cite}

\ifx\pdfoutput\undefined
\usepackage{graphicx}
\else
\usepackage[pdftex]{graphicx}
\fi

\usepackage{float} 
\usepackage[font=small]{caption}
\usepackage{subcaption}
\usepackage{xcolor}

\begin{document}

\author[1]{Oleg O.~Novikov\thanks{o.novikov@spbu.ru}}

\author[1,2]{Andrey A.~Shavrin\thanks{shavrin.andrey.cp@gmail.com}}

\affil[1]{Saint Petersburg State University, 7/9 Universitetskaya nab.\\
St. Petersburg, 199034, Russia}

\affil[2]{ITMO University, 49 Kronverksky Pr., St. Petersburg, 197101, Russia}

\title{Holographic model for the first order phase transition in the composite Higgs boson scenario}

\maketitle

\begin{abstract}
The composite Higgs model assumes that the Higgs field arises as the pseudo-Goldstone mode corresponding to a dynamical symmetry breaking in a new strongly coupled sector. We present a soft-wall holographic model where such symmetry breaking occurs as a first order phase transition. In this case the bubble nucleation in the early universe becomes possible. To study the homogeneous solutions in the models of this type we present the perturbation theory approach.
We estimate the gravitational wave spectrum produced during the nucleation phase and find it to be detectable with the planned gravitational wave detectors.
\end{abstract}

\section{Introduction}

The Standard model with minimal Higgs sector continues to be in the good agreement with LHC observations, however it can not address a number of astrophysical and cosmological observations. The nature of the Dark matter remains to be mystery. As its contribution to the energy density of the matter is approximately five times the contribution of the known particles, this shortcoming may be seen as one of the most pressing problem of the modern fundamental physics. Another issue arising from the cosmological considerations is the baryonic asymmetry of the universe, i.e. the prevalence of the matter of the antimatter. According to Sakharov\cite{Sakharov:1967dj,Sakharov:1991,White:2016nbo}, to explain such asymmetry three conditions should be satisfied simultaneously:
\begin{itemize}
    \item The baryonic number must not be conserved. While this symmetry is conserved in the perturbative Standard model, its violation become possible through the nonperturbative sphaleron processes at the electroweak phase transition temperatures\cite{Kuzmin:1985mm,Shaposhnikov:1986jp,Arnold:1987mh,Klinkhamer:1984di,Michael:2003}. Because the baryonic number conservation originates from the accidental symmetry its perturbative violation is common in the Standard model extensions but they are strongly constrained, primarily by the proton decay searches.
    \item The $\mathcal{C}$ and $\mathcal{CP}$ conservation must be broken.
    This is realized in the Standard model through the complex phases in the Cabibbo-Kobayashi-Maskawa (CKM) and Pontecorvo-Maki-Nakagawa-Sagata (PMNS) mixing matrices. Nevertheless, this violation is considered to be too small to account for the observed baryonic asymmetry. The constraints on the new sources of the $\mathcal{C}$ and $\mathcal{CP}$ violations come both from the collider searches and from the experiments with atoms and molecules \cite{khriplovich2012cp,ACME:18,chen2021heavy,zakharova2022rotating}.
    \item The thermal equilibrium must be violated. This may occur during the first-order phase transition that allow bubble nucleation to occur \cite{Zeldovich:411756,PhysRevD.15.2929}. The attractive feature of the Standard model is that, as we mentioned before, the baryonic violation becomes significant during the electroweak phase transition. Another important signature of such process would be a gravitational wave production during the nucleation phase \cite{kosowsky69s,turner1992wilczek,kosowsky1993gravitational,kamionkowski1994gravitational,caprini2016science,Cai:2017tmh,Weir:2017wfa,Caprini:2018mtu,Geller:2018mwu,Ellis:2020awk,caprini2020detecting,Bai:2021ibt}. However, theoretical studies of the thermal behaviour of the Higgs potential point to the crossover nature of such phase transition in the minimal Standard model. Thus, some modification of the Higgs sector or some unrelated first-order phase transition at higher temperatures is required\cite{Rummukainen:1998as,Kajantie:1996mn,Kajantie:1995kf}.
\end{itemize}

One of the popular extensions of the Standard model is the composite Higgs model that can alleviate to a certain degree the naturalness problems of the fundamental scalar Higgs field\cite{Contino:2003ve,Agashe:2004rs,Contino:2010rs,bellazzini2014composite,Panico:2015jxa}. In this type of the models the Higgs scalar is assumed to be a pseudo-Nambu-Goldstone particle originating from a dynamical symmetry breaking of the approximate global symmetry $\mathcal{G}$ to a subgroup $\mathcal{H}$ in a new strongly coupled sector, just as a pi-meson is a pseudo-Nambu-Goldstone boson associated with the breaking of the chiral symmetry in the quantum chromodynamics (QCD). Such models may be studied with help of the effective field theory approaches.
However, if such strongly coupled sector can be treated as a Yang-Mills gauge theory with sufficiently large number of colours $N_c$, the bottom-up holographic constructions can be applied. While the hard-wall models are easier to study, the AdS/QCD experience teaches us that the soft-wall models are more natural, with example of the soft-wall composite Higgs model constructed in \cite{Espriu:2017mlq,Katanaeva:2018frz,Espriu:2020hae,Falkowski:2008fz,Bellazzini:2014yua,Csaki:2022htl,Afonin:2022qkl,Elander:2020nyd,Elander:2021bmt,Elander:2023aow}.

Among other holographic composite Higgs constructions, we would like to mention the top-down inspired models in 
\cite{Erdmenger:2020lvq,Erdmenger:2020flu}.

The strong dynamics that results in the emergence of the composite Higgs may influence the $\mathcal{CP}$ violating  physics in important ways \cite{Chung:2021xhd,Chung:2021ekz,Cacciapaglia:2020jvj,Guan:2019qux}. It also may influence the nature of the electroweak phase transition \cite{Bruggisser:2018mrt,Bruggisser:2018mus}.
However, it also introduces novel phase transitions not present in the Standard model.

If we compare the composite Higgs model with QCD we can distinguish two phase transitions: the confinement-deconfinement phase transition and the $\mathcal{G}\rightarrow\mathcal{H}$ phase transition. In the holographic description the confinement-deconfinement transition is associated with the Hawking-Page phase transition\cite{Hawking:1982dh}. When the black hole geometries become thermodynamically preferred to the horizonless geometries, this affects the behavior of the long strings and, hence, the behavior of the long Wilson loops in the dual gauge theory. Therefore, the confinement-deconfinent phase transition involves the study of the gravitational or dilaton-gravitational dynamics. For example, in \cite{agashe2020cosmological,agashe2021phase} such dynamics was studied for the hard-wall model of the composite Higgs and the gravitational signatures of this phase transition were estimated. QCD induced phase transitions in the braneworld Randall–Sundrum models were studied with help of the dual holographic models in \cite{vonHarling:2017yew,Agrawal:2021alq,vonHarling:2023dfl}.

In this paper we study the $\mathcal{G}\rightarrow\mathcal{H}$ phase transition instead, which is analogous to the chiral phase transition in the QCD. The chiral phase transition was studied in the bottom-up AdS/QCD models through the dynamics of the scalar field in the asymptotically AdS spacetime with nonlinear potential that is dual to the chiral condensate \cite{cherman2009chiral,gherghetta2009chiral,guralnik2011dynamics,colangelo2012temperature,li2013dynamical,he2013phase,bartz2014dynamical,chelabi2016chiral,fang2016chiral,fang2016chiral2,bartz2016chiral,fang2019chiral}. In this paper we apply similar approach to construct the bottom-up holographic model of the composite Higgs admitting the first-order phase transition through the development of the condensate violating $\mathcal{G}$-symmetry. While in the previous AdS/QCD papers similar problem was treated numerically, we also apply the perturbation theory.

\section{The composite Higgs scenario}

The composite Higgs model assumes the new strong hypercolor gauge interaction between some fundamental fermions $\Psi_I$ or some other matter field with a mass gap of order $\mu_{IR}\sim 1-10\,{\rm TeV}$. $I$ denotes the index for some approximate hyperflavor symmetry $\mathcal{G}$ that is broken at low energies to its subgroup $\mathcal{H}$. The hypercolor number $N$ is assumed to be large. The Standard model fields (omitting Higgs field) are coupled through the gauging of the $U(1)_Y\times SU(2)_L$ subgroup in $\mathcal{H}$. The Lagrangian is given by,

\begin{equation}
    \mathcal{L}_{\rm tot}=\mathcal{L}_{\rm HC} + \mathcal{L}_{\rm SM}+ B_\mu J_Y^\mu+W_{\mu}^k J_L^{k,\mu}
    + \Big[\sum_{r}\bar{\psi}_r\mathcal{O}_r+\mathrm{h.c.}\Big],
\end{equation}
where $\mathcal{L}_{\rm HC}$ is the new strongly coupled sector consisting of the new fundamental matter fields $\Psi_I$ and their hypercolor interaction; $\mathcal{L}_{\rm SM}$ is the weakly coupled sector of the Standard model fields (excluding Higgs); $B_\mu$ and $W_{\mu}^k$ are $U(1)_Y$ and $SU(2)_L$ gauge fields respectively; $\psi_r$ are the fermions of the Standard model (left and right quarks and leptons), $J_Y^\mu$, $J^{k,\mu}$ are conserved currents in $\mathcal{L}_{\rm HC}$ associated with the $U(1)_Y$ and $SU(2)_L$ symmetries correspondingly; $\mathcal{O}_r$ are some composite operators from the hypercolor sector.

The models with different cosets $\mathcal{G}/\mathcal{H}$ are studied \cite{DaRold:2019ccj,Cheng:2020dum,Cacciapaglia:2019ixa,Espinosa:2011eu,Chala:2016ykx,Xie:2020bkl,Bian:2019kmg,Chala:2012af,Nevzorov:2015sha,Davighi:2018xwn,Frandsen:2023vhu,Fujikura:2023fbi}.
In the minimal variant the symmetry group $\mathcal{G}=SO(5)\times U(1)_{B-L}$ is broken to $\mathcal{H}=SO(4)\times U(1)_{B-L}\simeq SU(2)_L\times SU(2)_R\times U(1)_{B-L}$. $U(1)_Y$ arises as a subgroup of $SU(2)_R\times U(1)_{B-L}$. In the following we assume this scenario though our results may be easily generalized for a larger hyperflavor symmetry provided that $\mathcal{G}\to \mathcal{H}$ breaking happens in a sufficiently diagonal way in $\mathcal{G}/\mathcal{H}$.

The breaking of the symmetry at low energies may be associated with the development of the symmetry-violating condensate. We will assume that just like a chiral condensate in QCD it corresponds to the v.e.v. of the bilinear operator constructed from the hypercolored fermions,
\begin{equation}
\Sigma_{IJ}=\langle\bar{\Psi}_I\Psi_J\rangle,\label{CondensateOperator}
\end{equation}
where $I$ and $J$ are indices of the fundamental representation of $\mathcal{G}$.
This implies the non-anomalous conformal dimension $\Delta=3$.
Let us denote broken symmetry generators as $T_\alpha$. Then, if we neglect the approximate nature of the hyperflavor symmetry $\mathcal{G}$, at low energies the condensate experiences massless Goldstone fluctuations $\pi_\alpha$,
\begin{equation}
    \Sigma=\xi^T\Sigma_0\xi,\quad
    \Sigma_0=\begin{pmatrix}0_{4\times 4}&0\\0&\varsigma\end{pmatrix},
    \xi=\exp\Big(-i\pi_\alpha T_\alpha/f_\pi\Big),\label{CondensateAnsatz}
\end{equation}
where $f_\pi\sim\frac{\sqrt{N}}{4\pi}\mu_{IR}$ is analogous to $\pi$-meson decay constant in QCD.

However the interaction between Standard model and hypercolor sectors explicitly breaks the hyperflavor symmetry $\mathcal{G}$. As result, radiative processes produce the potential for $\pi_\alpha$. It is these pseudo-Goldstone fields that play the role of the Higgs field. The breaking of the electroweak symmetry that requires $\langle \pi_\alpha\rangle\neq 0$ is associated with the misalignment of the vacuum within $\mathcal{G}$ compared to $\mathcal{H}$. This is accompanied by the mixing between the elementary gauge bosons and fermions in $\mathcal{L}_{\rm SM}$ with the vector meson-like and baryon-like bound states in $\mathcal{L}_{\rm HC}$. The fermion mixing and the resulting masses are strongly depending on the anomalous dimension of the $\mathcal{O}_r$ operators, which may explain the hierarchy of the fermion masses.

While in the preceding discussion the only explicit breaking of $\mathcal{G}$ was coming from its interactions with the Standard model fields, $\mathcal{L}_\text{HC}$ itself may contain terms violating this flavour symmetry e.g. nondiagonal mass terms $m_{IJ}\bar{\Psi}_I\Psi_J+\mathrm{h.c.}$ In this paper we will neglect such contributions.

\section{Holographic model}

We assume that $\mathcal{L}_\text{HC}$ is a strongly coupled Yang-Mills theory with large number of colors $N$. Then we may employ AdS/CFT duality to describe it with help of the weakly coupled 5d theory with gravity. Taking the soft-wall bottom-up AdS/QCD model \cite{karch2006linear,erlich2005qcd,kwee2008pion,colangelo2008light} as an example, we consider the following model,
\begin{equation}\label{Holography:TotalLagrangian}
S_{\rm tot}=S_{\rm grav+\phi}+S_{\rm X}+S_{\rm A}+S_{\rm SM}.
\end{equation}
Here the first part is the Einstein-Dilaton action,
\begin{equation}
    S_{\rm grav+\phi}=\frac{1}{l_P^3}\int d^5x\sqrt{|g|}e^{2\phi}
    \Big[-R+2|\Lambda|-4g^{ab}\partial_a\phi\partial_b\phi-V_\phi(\phi)\Big],
\end{equation}
where $a,b=0,\ldots 4$
with $|\Lambda|=\frac{6}{L^2}$ and $(L/l_P)^4\sim N$. The second part is the action for the scalar field $X$ dual to $\Sigma$,
\begin{equation}
S_{\rm X}=\frac{1}{k_s}\int d^5x\sqrt{|g|}e^{\phi}
\Bigg[\frac{1}{2}g^{ab}\operatorname{Tr}\Big(\nabla_a X^T\nabla_b X\Big)-V_X(X) \Bigg],
\label{Xaction}
\end{equation}
where the scale $k_s$ is introduced to keep the dimensionality of $X$ similar to the dimensionality of the 4d scalar field. The choice would determine the normalization of the $X$-field and of the coefficients in the potential $V_X$. In our paper, we will take $k_s=L$. We will take the following potential that allows the symmetry breaking $\mathcal{G}\rightarrow\mathcal{H}$ with the mass adjusted to the conformal dimensionality of the dual operator,
\begin{equation}
V_X(X)=\operatorname{Tr}\Big(-\frac{3}{2L^2}X^TX-\frac{{v_4}}{4}(X^TX)^2+L^2\frac{{v_6}}{6}(X^TX)^3\Big)
\end{equation}
Where ${v_4}$ and ${v_6}$ are dimensionless.
For ${v_4}>0,{v_6}>0$ the phase transition will be the first order (whereas ${v_4}\leq 0$, ${v_6}\geq 0$ gives the second order).

The covariant derivative is,
\begin{equation}
\nabla_a X=\partial_a X+[A_a,X]
\end{equation}
where the $A_a$ gauges $\mathcal{G}$ flavor group and is dual to the current operators $J_{IJ}^\mu$ in the hypercolor sector. Its kinetic term is given by the third part of the action,
\begin{equation}
S_\text{A}=
-\frac{1}{g_5^2}\int d^5x\sqrt{|g|}e^\phi g^{ac}g^{bd}F_{ab}F_{cd}
\end{equation}
Finally, the interaction with the Standard model fields is given by the boundary term,
\begin{equation}
S_{\rm SM}=\epsilon^4\int_{z=\epsilon} d^4x\sqrt{|g^{(4)}|}\Bigg[\mathcal{L}_{\rm SM}
+c_Y B_\mu \operatorname{Tr}\Big(T_Y A^\mu\Big)
+c_W W_{k,\mu} \operatorname{Tr}\Big(T_k A^\mu\Big) +\mathcal{L}_{\psi} \Bigg]
\end{equation}
where $g^{(4)}_{\mu\nu}$ is the induced metric, $T_Y$ and $T_k$ are the generators of the electroweak group embedded into $\mathcal{G}$, and $\mathcal{L}_{\psi}$ is responsible for the interaction with the Standard model fermions not considered in this paper.

To study the phase transition at the finite temperature we make study this system on a space with Euclidean signature and periodic time coordinate $\tau\sim \tau+2\pi T^{-1}$. We take the fixed metric and dilaton background of a planar black hole in the asymptotically Euclidean AdS spacetime,
\begin{equation}
ds^2=\frac{L^2}{\tilde{z}^2}A(\tilde{z})^2\Big(f(\tilde{z})d\tau^2+\frac{d\tilde{z}^2}{f(\tilde{z})}+d\vec{x}^2\Big),\quad \phi=\phi(\tilde{z})
\end{equation}
where the $\tilde{z}\rightarrow 0$ limit gives AdS metric and the function $f(\tilde{z})$ has zero corresponding to the planar black hole horizon,
\begin{equation}
    f(z_H)=0,\:
    \frac{|f'(z_H)|}{4\pi}=T, \:
    f(\tilde{z})\underset{\tilde{z}\rightarrow 0}{\longrightarrow} 1,\:
    A(\tilde{z})\underset{\tilde{z}\rightarrow 0}{\longrightarrow} 1,
\end{equation}

The background must be a solution to the Einstein-dilaton equations of motion determined by the potential $V_\phi$.
However, to simplify our treatment we take the metric to be just a solution of the Einstein equations and the dilaton to be a standard quadratic ansatz providing the soft-wall infrared cutoff,
\begin{equation}
f=1-\frac{\tilde{z}^4}{z_H^4},\quad \phi=\tilde{\phi}_2 \tilde{z}^2,\quad z_H=\frac{1}{\pi T}.
\end{equation}

We employ the AdS/CFT correspondence \cite{Maldacena:1997re,Witten:1998qj,Aharony:1999ti} to define the generating functional of the hypercolor sector with help of the theory in AdS which in the limit $N\gg 1$ is assumed to be in the quasiclassical regime,
\begin{equation}
\mathcal{Z}_\text{HC}[J]=\mathcal{Z}_\text{AdS}[J]\sim\exp\Big(-S_E[J]\Big),\label{ZAdSCFT}
\end{equation}
where $S_E$ is the on-shell Euclidean action computed for solution with the asymptotic behaviour near the boundary determined by the currents $J$. For the scalar field $X$ we define it as,
\begin{equation}
L\cdot X_{IJ}\sim \frac{\sqrt{N}}{2\pi}J_{IJ}\tilde{z}+\frac{2\pi}{\sqrt{N}}\Sigma_{IJ}\tilde{z}^3+\ldots,\label{Xbc}
\end{equation}
where the factor $\frac{\sqrt{N}}{2\pi}$ comes from the appropriate scaling of the bilinear fermion operator \eqref{CondensateOperator} with $N$ \cite{cherman2009chiral}.

In QCD the chiral phase transition happens at temperatures close to the confinement-deconfinement transition temperatures. Therefore, the proper description of such transition must take into account the interplay between the Einstein-dilaton and scalar sectors. Similar situation may be expected for the $\mathcal{G}\rightarrow\mathcal{H}$ phase transition in the composite Higgs model. Nevertheless, in this paper we will decouple these transitions from each other by neglecting the dynamics of the Einstein-dilaton part and the backreaction of the scalar fields.
We will also neglect the impact of the gauge fields $A_a$. In the 4d QFT this would correspond to the zero chemical potential for the fermionic charges. We leave the investigation of the impact of our approximations to the future work.

\section{Phase transition description}

The equation for the $X$-field is (for the Euclidean signature),
\begin{equation}
\frac{1}{\sqrt{g}}e^{-\phi}\partial_a\Big(\sqrt{g}e^{\phi}g^{ab}\partial_bX_{IJ}\Big)+\frac{3}{L^2}X_{IJ}
+{v_4} X^{IJ}\operatorname{Tr}(X^TX)
-L^2{v_6} X^{IJ}\operatorname{Tr}(X^TX)^2.\label{Xeq1}
\end{equation}

The homogeneous solution for the scalar field is assumed to be in the form,
\begin{equation}
X=\begin{pmatrix}0_{4\times 4} & 0\\ 0 & \frac{\sqrt{3}}{\sqrt{{v_4}}}L^{-1}\chi(\tilde{z})\end{pmatrix},\label{XAnsatz}
\end{equation}
so that $\chi$ is dimensionless. The equation \eqref{Xeq1} then becomes,
\begin{equation}\label{PTDescription:BackgroundField}
\tilde{z}^5e^{-\phi(\tilde{z})}\partial_{\tilde{z}}\Big(\frac{1}{\tilde{z}^3}e^{\phi(\tilde{z})}f(\tilde{z})\partial_{\tilde{z}}\chi\Big)+3\chi+3\chi^3-\gamma\chi^5=0,
\end{equation}
where we introduced,
\begin{equation}
\gamma=9\frac{{v_6}}{{v_4}^2}.
\end{equation}
Near the anti-de Sitter boundary $\tilde{z}\rightarrow 0$ the solution behaves as,
\begin{equation}\label{asymptotic:border}
\chi\sim j\frac{\tilde{z}}{z_H}+\sigma \frac{\tilde{z}^3}{z_H^3}+\ldots.
\end{equation}
On the other hand, from the ansatz \eqref{CondensateAnsatz} and the boundary condition \eqref{Xbc} we have,
\begin{equation}
\chi\sim \frac{\sqrt{{v_4}}}{\sqrt{3}}\Big[\frac{\sqrt{N}}{2\pi}J\tilde{z}+\frac{2\pi}{\sqrt{N}}\varsigma \tilde{z}^3+\ldots\Big],
\end{equation}
where $J$ is the source for $\varsigma$. Comparing it with \eqref{asymptotic:border} we may identify our dimensionless constants with,
\begin{equation}
j=\frac{\sqrt{{v_4} N}}{2\pi\sqrt{3}}z_H J,\quad\sigma=\frac{2\pi\sqrt{{v_4}}}{\sqrt{3N}}z_H^3\varsigma.
\label{SigmaNorm}
\end{equation}
In this paper we will consider only the case $j=0$. Then the solution may be represented as a polynomial series in $\tilde{z}$ without $\tilde{z}^n\ln{\tilde{z}}$ terms.

On the other hand, near the horizon $\tilde{z}\rightarrow z_H$ the solution with the finite action behaves as,
\begin{equation}
\chi\sim \zeta_H+\omega\cdot\Big(1-\frac{\tilde{z}}{z_H}\Big)+\ldots,
\end{equation}
and is polynomial in $(1-\tilde{z}/z_H)$.

We rescale the coordinate $\tilde{z}=z_H z$. Then,
\begin{equation}\label{EoM:general}
z^5e^{-\phi_2z^2}\partial_{z}\Big(\frac{1}{z^3}e^{\phi_2z^2}\tilde{f}(z)\partial_{z}\chi\Big)+3\chi+3\chi^3-\gamma\chi^5=0,
\end{equation}
where $\tilde{f}(z)=1-z^4$,
and the only free parameter in this equation is,
\begin{equation}
\phi_2=\tilde{\phi}_2z_H^2,
\end{equation}
which is high for the low temperatures and low for the high temperatures.

Notice that while the ansatz \eqref{XAnsatz} manifestly violates the original global symmetry $\mathcal{G}$, the equation \eqref{EoM:general} has the symmetry under the transform $\chi\mapsto -\chi$. This reflection symmetry may be considered a residual of the original symmetry $\mathcal{G}$ coming from the rotation that changes the direction of the symmetry breaking to the opposite one. The only reflection symmetric solution is the trivial one $\chi=0$ which corresponds to the $\mathcal{G}$-symmetric $X$-field configuration. The phase transition  associated with $\mathcal{G}\to\mathcal{H}$ breaking is caused by the development of the nonzero condensate $\Sigma$ \eqref{CondensateAnsatz} that in the dual model corresponds to the nonzero $\chi$ solution non-invariant under the reflection transformation. Thus, the phase transition in this language is associated with the breaking of this reflection symmetry.

The free energy for \eqref{ZAdSCFT} is given by,
\begin{equation}
F=-T\ln{Z}=TS_E.
\end{equation}
As in this paper we neglect the all the dynamics except of the scalar field, we will replace the full $S_E$ with $S_{\text{X},E}$. Then for the homogeneous solutions,
\begin{equation}
F=\Big(\int d^3x\Big)\cdot\frac{6\pi}{{v_4}} \frac{1}{z_H^4} \mathcal{F},\label{FreeEnergyDim}
\end{equation}
where $\mathcal{F}$ is completely determined by the solution of \eqref{EoM:general} for given $\phi_2$ and $\gamma$,
\begin{equation}
\mathcal{F}=\int_0^1 dz\, \frac{1}{z^5}e^{\phi_2 z^2}\Big[\frac{z^2}{2}(\partial_z\chi)^2-\frac{3}{2}\chi^2-\frac{3}{4}\chi^4+\frac{\gamma}{6}\chi^6\Big],
\end{equation}
when $j=0$ one may integrate the derivative term by parts so that the boundary term vanishes. Then this expression can be simplified using \eqref{EoM:general} to,
\begin{equation}
\mathcal{F}=\int_0^1 dz\, \frac{1}{z^5}e^{\phi_2 z^2}\Big[\frac{3}{4}\chi^4-\frac{\gamma}{3}\chi^6\Big].
\end{equation}
From this representation of the free energy one may expect that for small $\chi$ the free energy increases whereas for larger $\chi$ it starts to decrease. This argument supported by the perturbation treatment below justifies our choice of the $V_X(X)$.

We study this equation numerically. Because $z=0$ and $z=1$ are singular points in this equation we approximate the function there with the series solutions and match them together using Runge-Kutta method in the intermediate region.
We also found that if the both series are obtained up to the 30th order, matching them at $z=0.5$ without any intermediate numerical solution leads to the acceptable error.

\begin{figure}[h]
    \centering
    \includegraphics[width=0.45\textwidth]{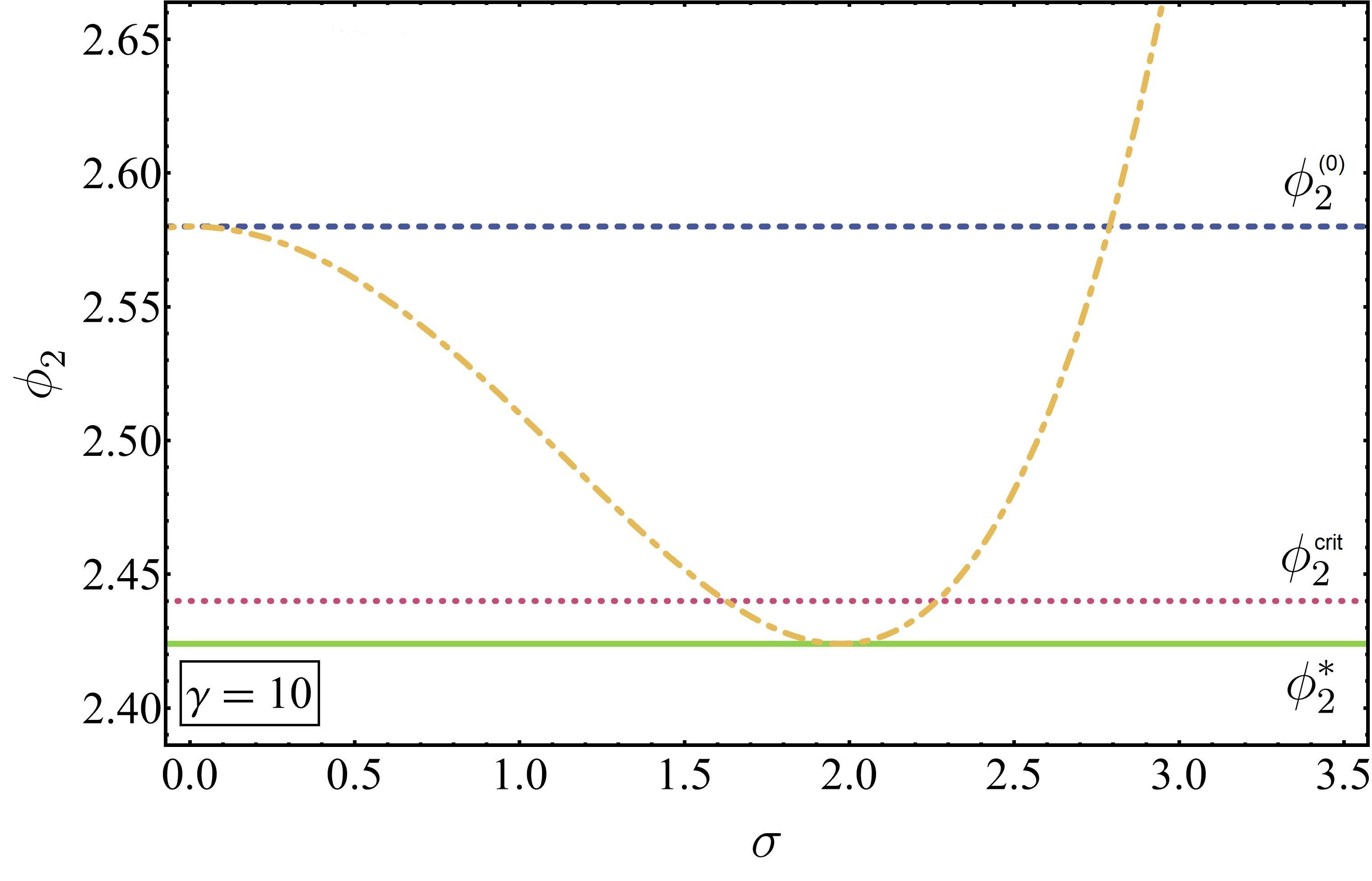}
    \captionsetup{justification=raggedright,singlelinecheck=false}
    \caption{The typical behavior of the phase diagram illustrated with $\gamma=10$ case}
    \label{fig:phasediag}
\end{figure}

\begin{figure}[h]
    \centering
    \includegraphics[width=0.45\textwidth]{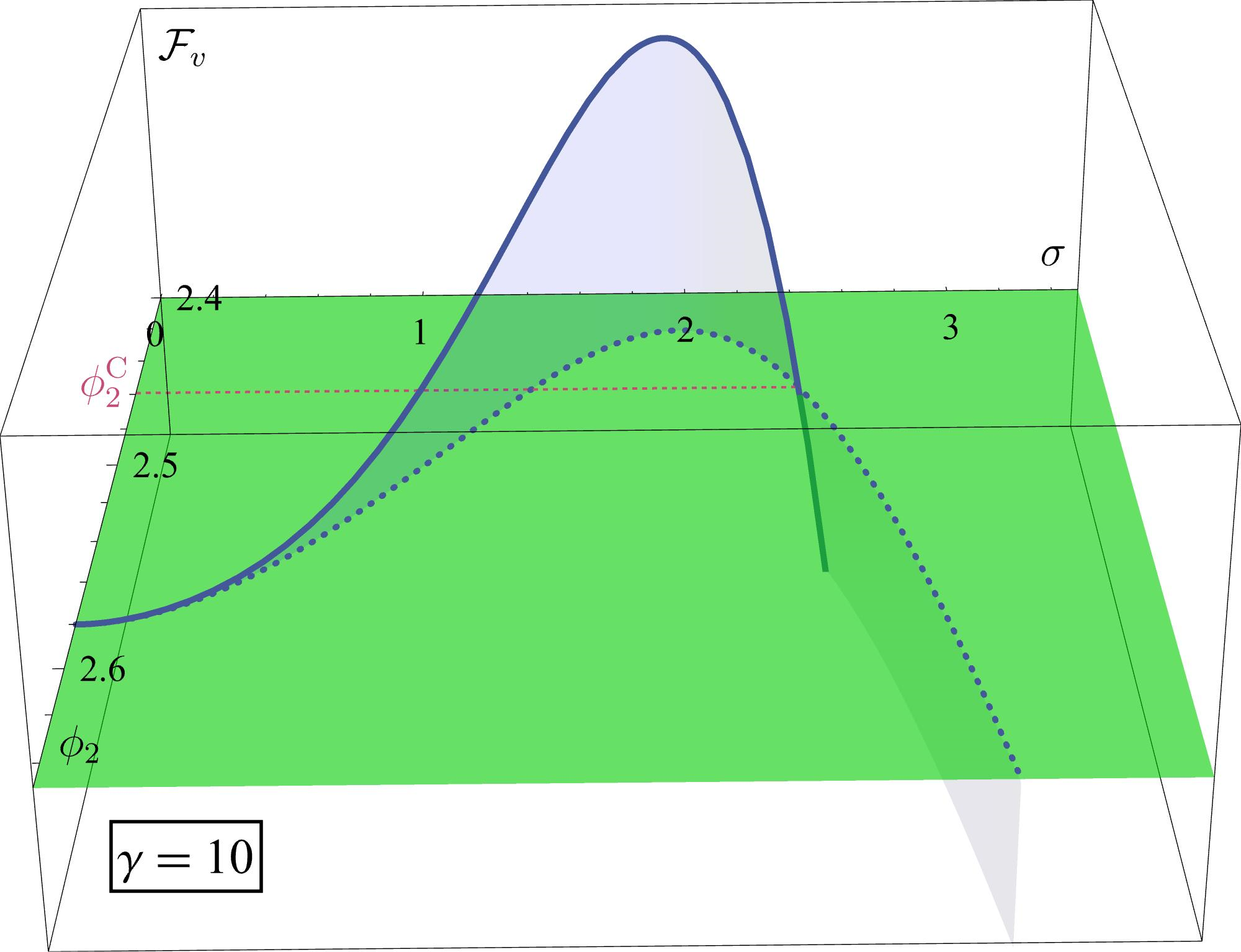}
    \captionsetup{justification=raggedright,singlelinecheck=false}
    \caption{The typical behavior for the free energy density illustrated with $\gamma=10$ case}
    \label{fig:FreeEnergy}
\end{figure}

The typical behavior of the numerical solutions in the model considered is depicted on the Fig. \ref{fig:phasediag} whereas the corresponding free energy is plotted on Fig. \ref{fig:FreeEnergy}. One may notice that the behavior is qualitatively similar to the Landau first phase transition model.

As one can see, at low values of $\phi_2<\phi_2^\ast$ (that corresponds to the high temperatures) no non-trivial solution exists. Therefore at high temperatures only the phase with $\sigma=0$ exists that is characterized by non-broken reflection and $\mathcal{G}$ symmetries. However at point $\phi_2=\phi_2^\ast$ two nontrivial solutions appear that diverge as two branches. We will call the solution belonging to the branch with larger $|\sigma|$ as the upper solution whereas the solution with the smaller $|\sigma|$ as the lower solution. At this point the trivial solution has the lowest free energy, thus, corresponding to the stable phase with $\sigma=0$, whereas the upper solution corresponds to the new metastable phase with $\sigma\neq 0$ and broken symmetry, while the lower solution is the unstable maximum of the barrier. At certain point $\phi_2=\phi_2^\text{crit}$ the free energy of the upper nontrivial solution becomes lower than the free energy of the trivial solution, while the lower nontrivial solution still has higher free energy. Because of the barrier represented by the lower solution, this represents the first order transition between the phase with $\sigma=0$ and phase with $\sigma\neq 0$. The lower solution  disappears when $\phi_2=\phi_2^{(0)}\simeq 2.58$, the value not depending on $\gamma$ parameter. From this point the metastable phase with $\sigma=0$ no longer exists and at low temperatures only the phase with $\sigma\neq 0$ represented by the upper solution is possible.

\section{Perturbation theory}

To get better understanding of the phase diagram we will study it with help of the perturbation theory. First, let us rescale the field,
\begin{equation}
\chi=\sqrt{\lambda}\psi,
\end{equation}
so that the equation \eqref{EoM:general} takes the form,
\begin{equation}
z^5e^{-\phi_2z^2}\partial_{z}\Big(\frac{1}{z^3}e^{\phi_2z^2}\tilde{f}(z)\partial_{z}\psi\Big)
+3\psi+3\lambda\psi^3-\gamma\lambda^2\psi^5=0.\label{EoM:rescaled}
\end{equation}
Taking $\lambda$ to be a small parameter, we will represent the solution and parameter $\phi_2$ as,
\begin{align}
\psi&=\psi^{(0)}+\lambda\psi^{(1)}+\lambda^2\psi^{(2)}+\ldots,\\
\phi_2&=\phi_2^{(0)}+\lambda \phi_2^{(1)}+\lambda^2\phi_2^{(2)}+\ldots.
\end{align}
If we take,
\begin{equation}
\psi^{(0)}(1)=1,\quad \psi^{(n>0)}(1)=0,
\end{equation}
then $\lambda=\chi^2(1)=\zeta_H^2$. From Fig. \ref{fig:phasediag} one may expect that our perturbation theory should yield the solutions in the vicinity of the point that we already denoted as $\phi_2^{(0)}$.

For the zeroth order the equation \eqref{EoM:rescaled} is linear,
\begin{equation}
z^5e^{-\phi_2^{(0)}z^2}\partial_{z}\Big(\frac{1}{z^3}e^{\phi_2^{(0)}z^2}\tilde{f}(z)\partial_{z}\psi\Big)+3\psi^{(0)}=0,\label{EoM:0order}
\end{equation}
which we will rewrite in the form,
\begin{equation}
\mathcal{L}_0\psi^{(0)}=0.
\end{equation}
We may note that for the inner product,
\begin{equation}
(\xi,\eta)\equiv \int_0^1 dz\,\frac{1}{z^5}e^{\phi_2^{(0)}z^2}\xi(z)\eta(z),
\end{equation}
this operator is symmetric,
\begin{equation}
(\mathcal{L}_0\xi,\eta)=(\xi,\mathcal{L}_0\eta).
\end{equation}
The solutions that have the finite norm $(\psi^{(0)},\psi^{(0)})$ must satisfy the boundary conditions that we imposed in the previous section,
\begin{align}
&z\rightarrow 0,\quad\psi^{(0)}\sim \sigma^{(0)}\cdot z^3+\ldots,\\
&z\rightarrow 1,\quad \psi^{(0)}\sim 1+\omega^{(0)}\cdot (1-z)+\ldots.
\end{align}
The equation \eqref{EoM:0order} is basically a one-dimensional Schr\"{o}dinger equation with a potential depending on the parameter $\phi_2^{(0)}$. The solution with finite norm exists only for a single value $\phi_2^{(0)}\simeq 2.58$ that agrees with numerical computations in the previous section. It has $\sigma^{(0)}\simeq 4.41$. Regretfully, it seems that there is no analytic expression of $\psi^{(0)}$, so the perturbation theory has to be done in a semi-analytic fashion, treating $\psi^{(0)}$ as a new special function.

One can also construct the linearly independent solution $\tilde{\psi}^{(0)}$ by using their Wronskian,
\begin{equation}
\mathcal{W}=\psi^{(0)} \partial_z\tilde{\psi}^{(0)} - \tilde{\psi}^{(0)} \partial_z \psi^{(0)}=\frac{z^3}{\tilde{f}(z)}e^{-\phi_2^{(0)}z^2},
\end{equation}
which is non-normalizable,
\begin{align}
&z\rightarrow 0,\quad\tilde{\psi}^{(0)}\sim \frac{1}{2\sigma^{(0)}}\cdot (z+z^3\ln{z}+\ldots),\\
&z\rightarrow 1,\quad \tilde{\psi}^{(0)}\sim -\frac{1}{4}e^{-\phi_2^{(0)}}\ln(1-z)+\ldots.
\end{align}

In the first order only the cubic term plays the role,
\begin{equation}
\mathcal{L}_0\psi^{(1)}=\mathcal{G}_1,\:\: \mathcal{G}_1=-2\phi_2^{(1)}z^3\tilde{f}(z)\partial_z\psi^{(0)}-3(\psi^{(0)})^3.
\end{equation}
As is commonly done in the quantum mechanical perturbation theory, the value of $\phi_2^{(1)}$ may be obtained from the requirement of the finiteness of the norm of $\psi^{(1)}$ with help of the self-adjointness of $\mathcal{L}_0$,
\begin{equation}
(\psi^{(0)},\mathcal{L}_0\psi^{(1)})=(\mathcal{L}_0\psi^{(0)},\psi^{(1)})=0,
\end{equation}
from which we get,
\begin{equation}
\phi_2^{(1)}=-\frac{3}{2}\cdot\frac{\Big(\psi^{(0)},(\psi^{(0)})^3\Big)}{\Big(\psi^{(0)},z^3\tilde{f}(z)\partial_z\psi^{(0)}\Big)}.\label{PT:phi2_1}
\end{equation}
The solution for $\psi^{(1)}$ may be obtained by the variation of constants,
\begin{equation}
\psi^{(1)}=C^{(1)}(z)\psi^{(0)}+\tilde{C}^{(1)}(z)\tilde{\psi}^{(0)},
\end{equation}
where,
\begin{equation}
C^{(1)}(z)=\int_z^1 dz\,\frac{\mathcal{G}_1}{\mathcal{W}}\tilde{\psi}^{(1)},\quad \tilde{C}^{(1)}=\int_0^z dz\,\frac{\mathcal{G}_1}{\mathcal{W}}\psi^{(1)}.
\end{equation}
The condition \eqref{PT:phi2_1} together with this choice of the integration limits results in,
\begin{align}
&z\rightarrow 0,\quad \psi^{(1)}\sim \sigma^{(1)}\cdot z^3+\ldots,\\
&z\rightarrow 1,\quad \psi^{(0)}\sim \omega^{(1)}\cdot (1-z)\ldots.
\end{align}

Next orders of $\psi^{(n)}$ satisfy the equations,
\begin{equation}
\mathcal{L}_0\psi^{(n)}=\mathcal{G}_n,
\end{equation}
and can be obtained in the same fashion using $\mathcal{G}_n$ instead $\mathcal{G}_1$. In the second order the dependence on $\gamma$ appears,
\begin{equation}
\mathcal{G}_2=-2z^3\tilde{f}(z)\Big(\phi_2^{(2)}\partial_z\psi^{(0)}+\phi_2^{(1)}\partial_z\psi^{(1)}\Big)
-9(\psi^{(0)})^2\psi^{(1)}+\gamma(\psi^{(0)})^5,
\end{equation}
as result the second order correction to $\phi_2$ is a linear function on $\gamma$,
\begin{equation}
\phi_2^{(2)}=\phi_2^{(2),0}+\gamma\phi_2^{(2),\gamma}
\end{equation}
\begin{equation}
\phi_2^{(2),0}=-\frac{1}{2}\cdot\frac{\Big(\psi^{(0)},2z^3\tilde{f}(z)\partial_z\psi^{(1)}+9(\psi^{(0)})^2\psi^{(1)}\Big)}{\Big(\psi^{(0)},z^3\tilde{f}(z)\partial_z\psi^{(0)}\Big)},
\end{equation}
\begin{equation}
\phi_2^{(2),\gamma}=\frac{1}{2}\cdot\frac{\Big(\psi^{(0)},(\psi^{(0)})^5\Big)}{\Big(\psi^{(0)},z^3\tilde{f}(z)\partial_z\psi^{(0)}\Big)}.
\end{equation}

This way we obtained,
\begin{align}\label{Analytical:SigmaAndDilaton}
\sigma&\simeq \sqrt{\lambda}\Big(4.41-3.43\lambda+(2.52+0.95\gamma)\lambda^2\Big),\\ \phi_2&\simeq 2.58-1.68\lambda+(0.94+0.39\gamma)\lambda^2.\label{PT:results}
\end{align}
The point $\phi_2^\ast$ on Fig. \ref{fig:phasediag} when the symmetry-breaking phase and the barrier appear can be obtained as,
\begin{equation}
\frac{d\phi_2}{d\lambda}(\lambda_\ast)=0,\quad \lambda_\ast\simeq-\frac{\phi_2^{(1)}}{2(\phi_2^{(2),0}+\gamma\phi_2^{(2),\gamma})}.
\end{equation}
Obviously the second-order perturbation theory becomes useless for study of the phase diagram when $\lambda_\ast\gtrsim \frac{1}{2}$. For the values we obtained this requires $\gamma\gtrsim 2$.

The free energy may be represented as,
\begin{multline}
\mathcal{F}=\lambda^2\int_0^1dz\,\frac{1}{z^5}e^{\phi_2^{(0)}z^2}\Big[\frac{3(\psi^{(0)})^4}{4}\Big]
\\+\lambda^3\int_0^1dz\,\frac{1}{z^5}e^{\phi_2^{(0)}z^2}
\Big[\phi_2^{(1)}\frac{3(\psi^{(0)})^4}{4}+3(\psi^{(0)})^3\psi^{(1)}-\gamma\frac{(\psi^{(0)})^6}{3}\Big],
\end{multline}
which for the obtained solutions equal to,
\begin{equation}\label{Analytical:FreeEnergyDensity}
\mathcal{F}=2.18\lambda^2+(-1.27-0.69\gamma)\lambda^3.
\end{equation}
This gives us the critical point of the first-order phase transition (when the symmetry breaking phase has the same free energy as the symmetry preserving one),
\begin{equation}
\lambda_{\text{crit}}=\frac{2.18}{1.27+0.69\gamma}.
\end{equation}

\section{Modified perturbation theory}

There is a way to reorganize the perturbation series to cut the computations. Let us assume that $\gamma$ parameter is sufficiently large, so that,
\begin{equation}
\gamma=\frac{g}{\lambda},\quad g\sim 1.
\end{equation}
Then the $\chi^5$ term affects already the first order,
\begin{equation}
\mathcal{G}_1^{\text{mod}}=-2\phi_2^{(1)}z^3\tilde{f}(z)\partial_z\psi^{(0)}-3(\psi^{(0)})^3+g(\psi^{(0)})^3.
\end{equation}
The first order correction is then,
\begin{equation}
\phi_2^{(1)}=-\frac{3}{2}\cdot\frac{\Big(\psi^{(0)},(\psi^{(0)})^3\Big)}{\Big(\psi^{(0)},z^3\tilde{f}(z)\partial_z\psi^{(0)}\Big)}
+\frac{g}{2}\cdot\frac{\Big(\psi^{(0)},(\psi^{(0)})^5\Big)}{\Big(\psi^{(0)},z^3\tilde{f}(z)\partial_z\psi^{(0)}\Big)}.
\end{equation}
Not surprisingly, we get the result,
\begin{equation}
\sigma\simeq \sqrt{\lambda}\Big(4.41-3.43\lambda+0.95g\lambda\Big)
=\sqrt{\lambda}\Big(4.41-3.43\lambda+0.95\gamma\lambda^2\Big),
\end{equation}
\begin{equation}
\phi_2\simeq 2.58-1.68\lambda+0.39g\lambda
=2.58-1.68\lambda+0.39\gamma\lambda^2,
\end{equation}
that is consistent with \eqref{PT:results} when $\gamma$ is large. The advantage of this approach is that to obtain $\phi_2^{(1)}$ and make rough estimates on the critical values of $\phi_2$ and $\lambda$, one does not need to compute $\psi^{(1)}$.
Of course $\sigma^{(1)}$ correction is, as a matter of fact, important to lessen the errors at high $\lambda$.

\begin{figure}[h]
    \centering
    \subfloat[][]{\includegraphics[width=0.49\linewidth]{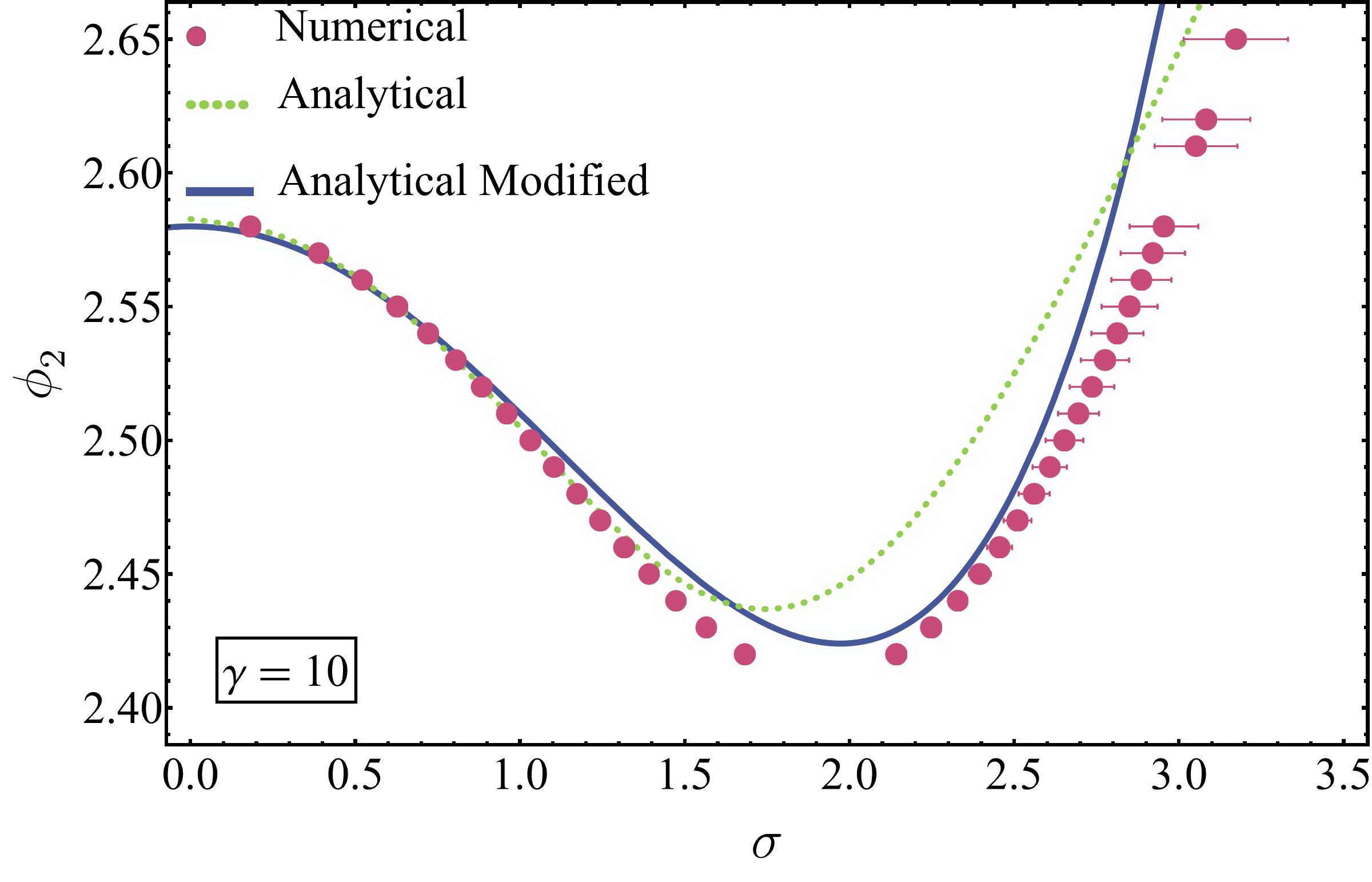}}
    ~
    \subfloat[][]{\includegraphics[width=0.49\linewidth]{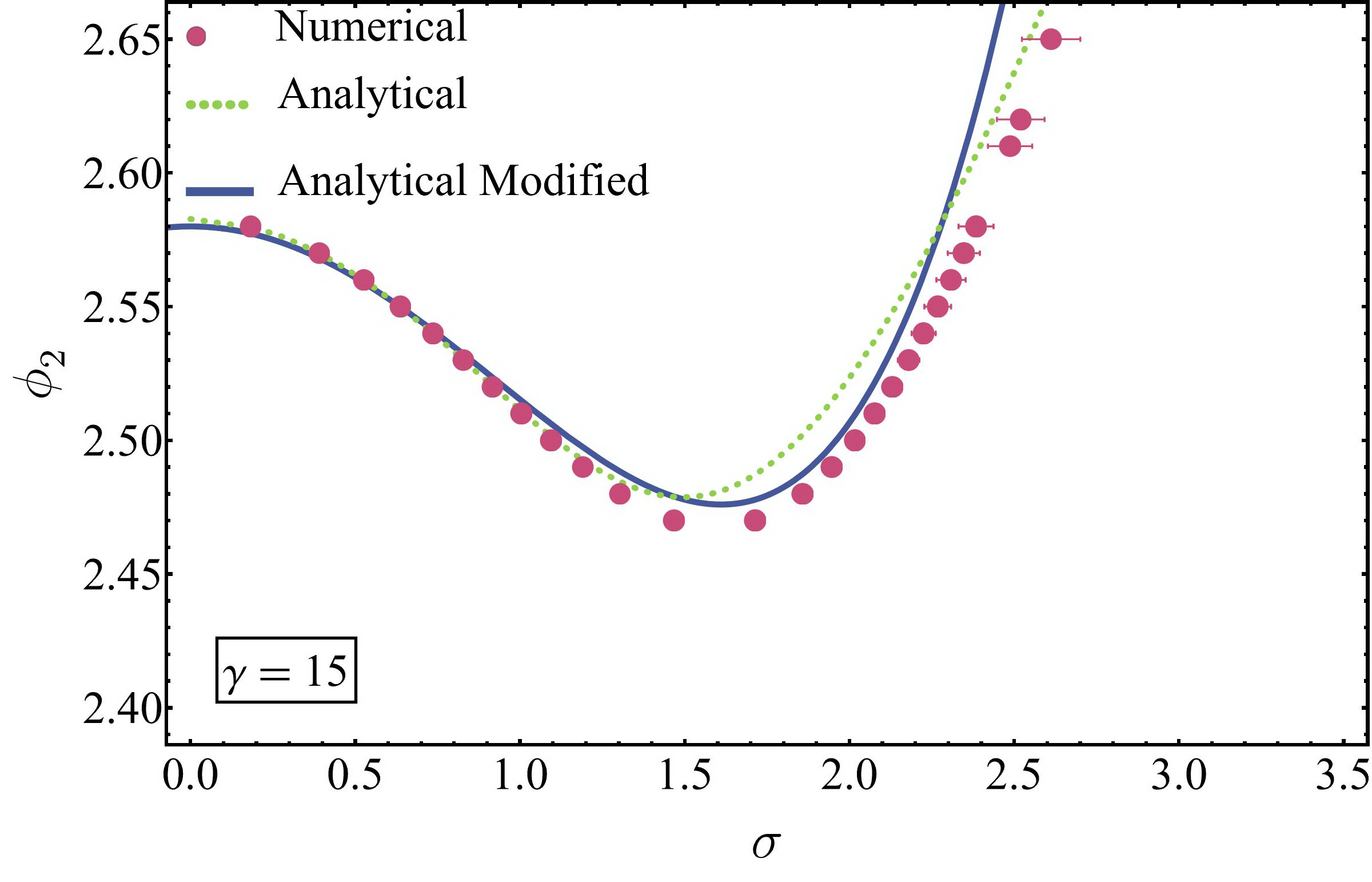}}
    \vspace{0.5cm}
    
    \subfloat[][]{\includegraphics[width=0.49\linewidth]{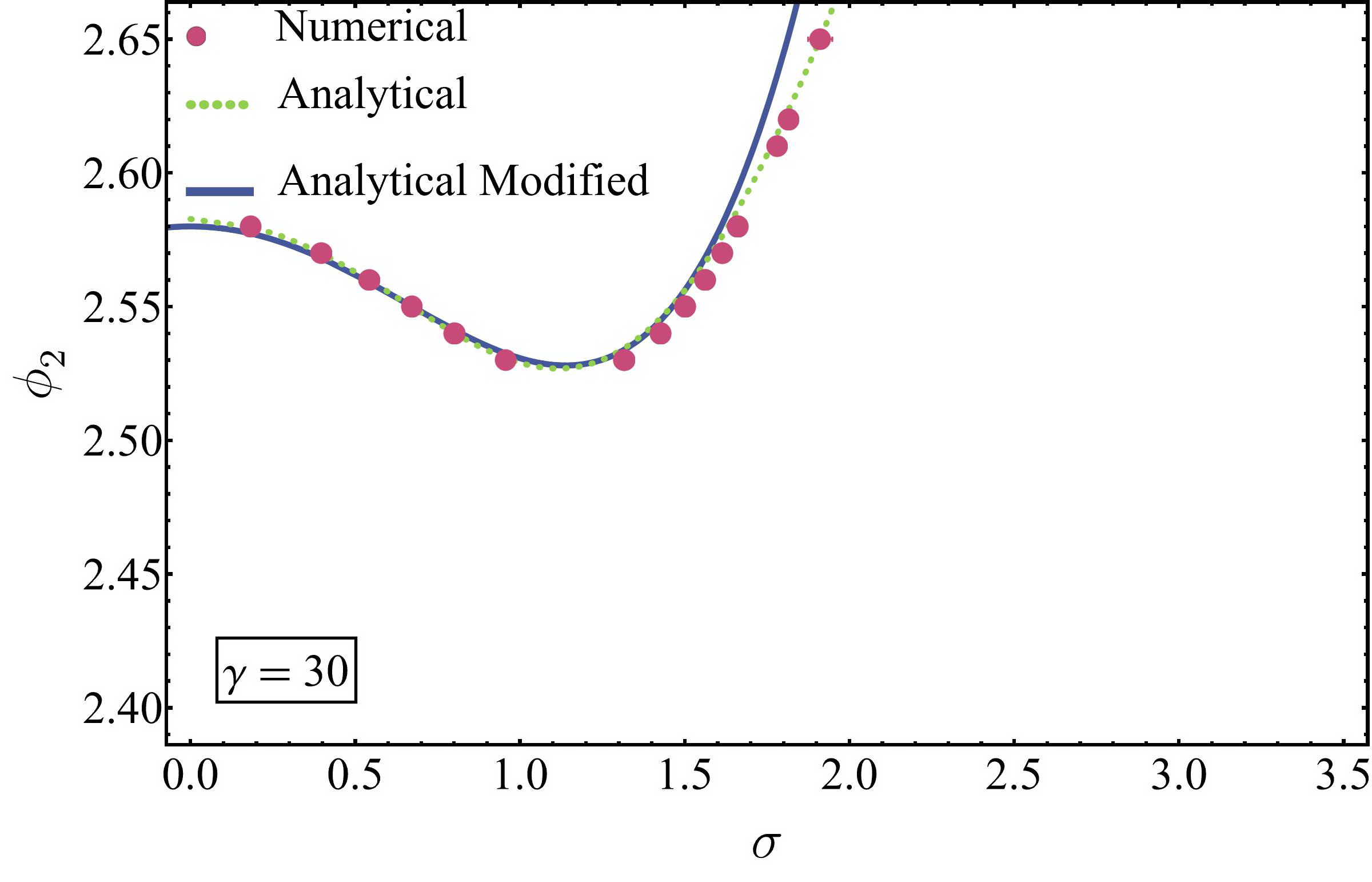}}
    ~
    \subfloat[][]{\includegraphics[width=0.49\linewidth]{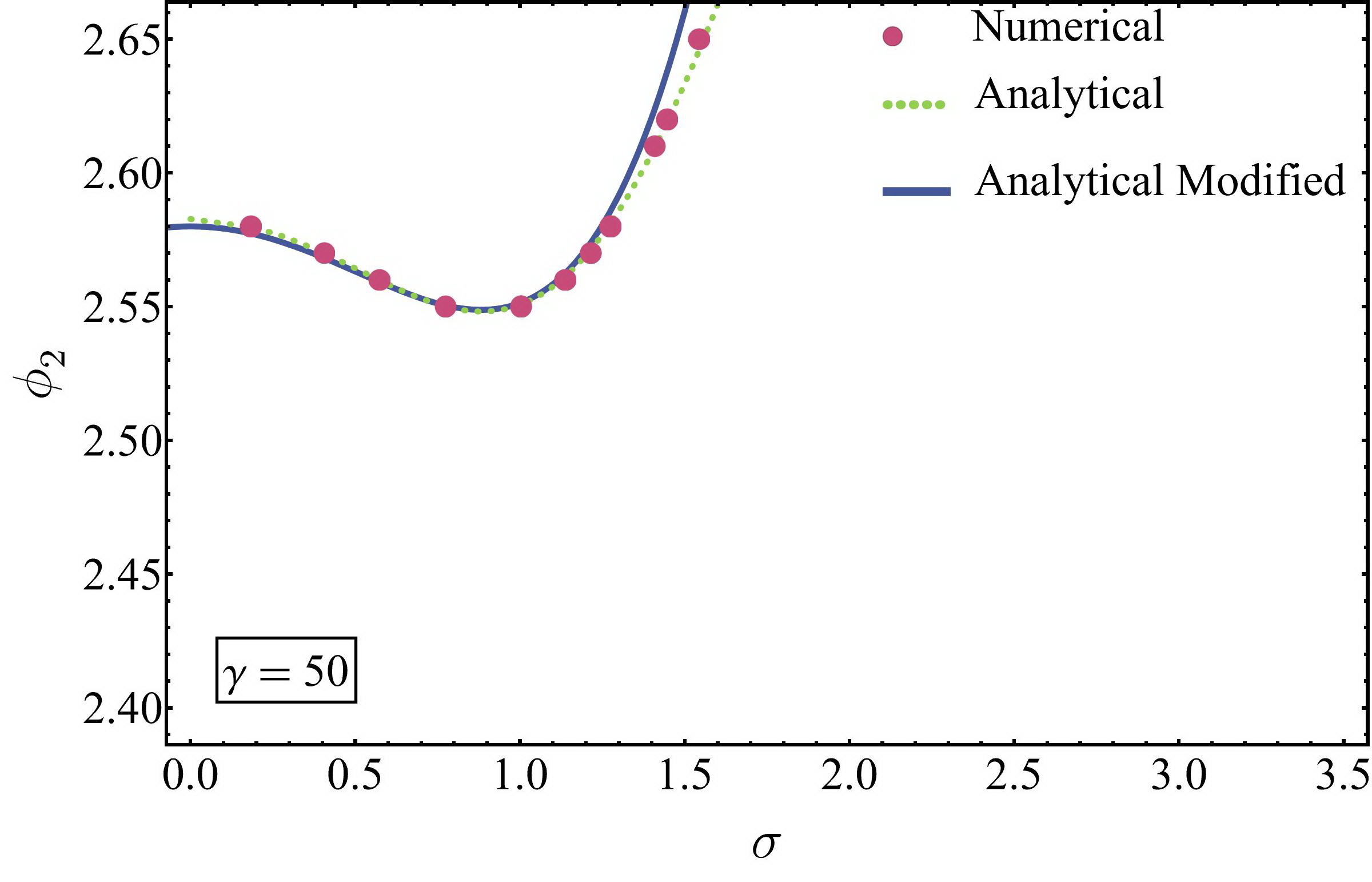}}
    
    \captionsetup{justification=raggedright,singlelinecheck=false}
    \caption{Comparison between the numerical computations, the second order of the perturbation theory and the first order of the modified perturbation theory (a) $\gamma=10$ (b) $\gamma=15$ (c) $\gamma=30$ (d) $\gamma=50$}
    \label{fig:comparison}
\end{figure}

The comparison between the perturbation theory and numerical computations is shown on the Fig. \ref{fig:comparison}. One may notice that the perturbation theory is successful in at least qualitative description of the phase diagram.

\section{Quantum effective potential for the condensate}

The bubble nucleation rate is determined by the free energy of the critical bubble,
the Arrhenius equation
\cite{Linde:1981zj,Rubakov:2017xzr,Hindmarsh:2020hop},
\begin{equation}\label{NucleationRate}
    \Gamma_\text{nucleation}
    = T^4 \left(\frac{F_\text{bubble}(R_\text{(crit)})}{2\pi T}\right)^\frac{3}{2}
    \exp\left(-\frac{F_\text{bubble}(R_\text{(crit)})}{T}\right).
\end{equation}
The bubble is assumed to be in the thin wall approximation, so that for its radius $R$ the free energy equals,
\begin{equation}
F_\text{bubble}(R)=4\pi R^2\tilde{\mu}-\frac{4\pi}{3}R^3 \Big(\frac{dF_{out}}{dV}-\frac{dF_{in}}{dV}\Big),
\end{equation}
where $dF_{out}/dV$ and $dF_{in}/dV$ are the free energy densities outside and inside the bubble respectively, and $\tilde{\mu}$ is the surface free energy density of the bubble. The critical radius for the bubble is determined by,
\begin{equation}
\frac{dF_\text{bubble}}{dR}\Big|_{R=R_\text{crit}} = 0.
\end{equation}

To obtain $\tilde{\mu}$ it will be useful to introduce the quantum effective action \cite{Kiritsis:2012ma}. Let us quickly elucidate this. First we define,
\begin{equation}
e^{-W[J]}\equiv\mathcal{Z}[J].
\end{equation}
The v.e.v. of a certain field $\Psi$ (also known as the `classical field')
associated with the current $J$ in the presence of the nonzero current may be obtained as,
\begin{equation}
\Psi_c(x)\equiv\langle\Psi(x)\rangle = \frac{\delta W[J]}{\delta J(x)}.
\end{equation}
Then the Legendre transform will give us the quantum effective action defined as,
\begin{equation}
\Gamma[\Psi_c]=W[J]-\int d^4x\, \Psi_c(x)J(x).
\end{equation}
The first variational derivative of the $\Gamma$ gives the current,
\begin{equation}
\frac{\delta\Gamma[\varsigma_c]}{\delta\Psi_c(x)}=-J(x).
\end{equation}
For $J=0$ we may obtain that,
\begin{equation}
\Psi_c=\mathrm{const}\cdot\varsigma
\end{equation}

On the homogeneous configurations the quantum effective action reduces to the quantum effective potential,
\begin{equation}
\Gamma = -\Big(\int d^4x\Big) \cdot U_{\text{eff}}
\end{equation}
One may notice that this effective potential is equal to the free energy density,
\begin{equation}
\Big(\int d^3x\Big)\cdot U_{\text{eff}}=F,
\end{equation}
therefore, using \eqref{FreeEnergyDim} we get in our case,
\begin{equation}
U_{\text{eff}}=\frac{6\pi}{{v_4}}\frac{1}{z_H^4}\mathcal{F}.\label{EffectivePotential}
\end{equation}
Our solutions with $J\sim j=0$ represent the extrema of this potential. In principle if we consider the solutions with $j\neq 0$ we may restore the full shape of the potential. However, one should note that if $J,j\neq 0$ then $\mathcal{F}$ diverges at $z\rightarrow 0$. To solve this one should renormalize the action by addition of the boundary term at $z=\epsilon$ by adding a boundary term as described in \cite{Skenderis:2002wp}.

As it is not needed for our solutions, we will not consider the renormalization of the homogeneous effective potential here. Instead we will assume that using the data about its extrema the renormalized potential may be extrapolated with,
\begin{equation}
        U_\text{eff} \simeq a_0 + a_2 \varsigma_c^2 + a_4 \varsigma_c^4 + a_6 \varsigma_c^6.
\end{equation}
For the data obtained in the previous sections we get $a_0\approx 0$ which ensures that the extrapolated free energy of the symmetry-preserving phase is close to zero.

To get the normalization factor for the fields with the canonical kinetic term we consider the small perturbations near the homogeneous solution. We will for simplicity consider only the perturbation of the nontrivial component,
\begin{equation}
X_{IJ}=\begin{pmatrix}0_{4\times 4}&0\\0&\frac{\sqrt{3}}{\sqrt{{v_4}}}L^{-1}\chi(\tilde{z})+\rho(\tau,\vec{x},\tilde{z})\end{pmatrix},
\end{equation}
where the perturbation has the asymptotic form,
\begin{equation}
L\cdot \rho\sim \frac{\sqrt{N}}{2\pi}\tilde{z}\Bigg[1-\frac{\tilde{z}^2}{2z_H^2}\ln\frac{\tilde{z}}{z_H}\Big(z_H^2\partial_\mu\partial^\mu+2\phi_2\Big)\Bigg]\mathcal{J}(\tau,\vec{x})
+\frac{2\pi}{\sqrt{N}}\delta\varsigma(\tau,\vec{x})\tilde{z}^3+\ldots,
\end{equation}
The contribution of the perturbations to the $W[J]$ may be obtained as,
\begin{equation}
W_{\rho}[\mathcal{J}]=W_{\rho}^{(bare)}[\mathcal{J}]+W_\rho^{(c-t)}[\mathcal{J}],
\end{equation}
where $W_{\rho}^{(bare)}$ is given by the bare action \eqref{Xaction} regularized by restricting integration domain to $\tilde{z}>\epsilon$,
\begin{equation}
W_\rho^{(bare)}=
\frac{1}{L}\int d^4x\int_\epsilon^{z_H} d\tilde{z}\sqrt{|g|}e^{\phi}
\Bigg[\frac{1}{2}g^{ab}\partial_a \rho\partial_b \rho-\frac{3}{2L^2}\rho^2+V_\rho(z)\rho^2\Bigg],
\end{equation}
where the potential,
\begin{equation}
V_\rho(z)=\frac{27}{2L^2}\xi^2-\frac{5}{2L^2}\xi^4\sim  \frac{9}{2L^2}\varsigma^2\tilde{z}^6
\end{equation}
plays little role near the boundary $\tilde{z}\rightarrow 0$ but important for the dependence of $\varsigma$ on $\mathcal{J}$. Using the equation of motion this bare action may be reduced to the boundary term,
\begin{equation}
W_{\rho}^{(bare)}=\frac{L^2}{2}\frac{e^{\phi}}{\tilde{z}^3}f(\tilde{z})\int d^4x\rho\partial_{\tilde{z}}\rho\Big|_{\tilde{z}=\epsilon},
\end{equation}
which or $\mathcal{J}\neq 0$ is divergent. As mentioned above, to deal with this divergence we introduce the boundary counterterm,
\begin{equation}
W_{\rho}^{(c-t)}=-\frac{L^2}{2}\int d^4x \rho(\tau,\vec{x},\epsilon)\Bigg[\frac{1}{\epsilon^4}
-\frac{1}{\epsilon^2z_H^2}\ln\frac{\epsilon}{z_H}\Big(z_H^2\partial_\mu\partial^\mu+2\phi_2\Big)\Bigg]\rho(\tau,\vec{x},\epsilon).
\end{equation}
In the limit $\epsilon\rightarrow 0$ the full contribution of the fluctuations to $W[J]$ becomes,
\begin{equation}
W_\rho[\mathcal{J}]=\int d^4x \Big(-\frac{N}{16\pi^2}\mathcal{J}\partial_\mu\partial^\mu\mathcal{J}+\mathcal{J}\delta\varsigma\Big),
\end{equation}
From this we get the classical field variable,
\begin{equation}
\delta\Psi_c=\langle\delta\Psi\rangle=\frac{\delta W_\rho}{\delta\mathcal{J}}=\delta\varsigma+\mathcal{J}\frac{\partial(\delta\varsigma)}{\partial\mathcal{J}}-\frac{N}{8\pi^2}\partial_\mu\partial^\mu\mathcal{J},
\end{equation}
so that at $\mathcal{J}=0$ we have $\delta\Psi_c=\delta\varsigma$. When $\frac{\partial(\delta\varsigma)}{\partial\mathcal{J}}$ may be neglected the resulting effective action would be of a free CFT. However this term reflects the conformal symmetry breaking by the background. As the equation on $\rho$ is linear then we may express $\delta\varsigma$ as,
\begin{equation}
\delta\varsigma\sim \frac{N}{4\pi^2}C\frac{1}{z_H^2}\mathcal{J}
\end{equation}
where $\phi_2$, ${v_4}$ and ${v_6}$ dependent coefficient $C$ is of order $1$.

Let us restrict ourselves to small momenta. Then the current variable is approximately,
\begin{equation}
\mathcal{J}\simeq \frac{2\pi^2z_H^2}{CN}\Big[1+\frac{1}{4C}\partial_\mu\partial^\mu z_H^2\Big]\Psi_c,
\end{equation}

The effective action then takes the form,
\begin{equation}
\Gamma_\rho[\delta\Psi_c] = \frac{\pi^2z_H^4}{4C^2N}
\int d^4x \Big(\partial_\mu\delta\Psi_c\partial^\mu\delta\Psi_c-\frac{4C}{z_H^2}(\delta\Psi_c)^2\Big).
\end{equation}

\begin{figure}[h]
    \centering
    \includegraphics[width=0.45\textwidth]{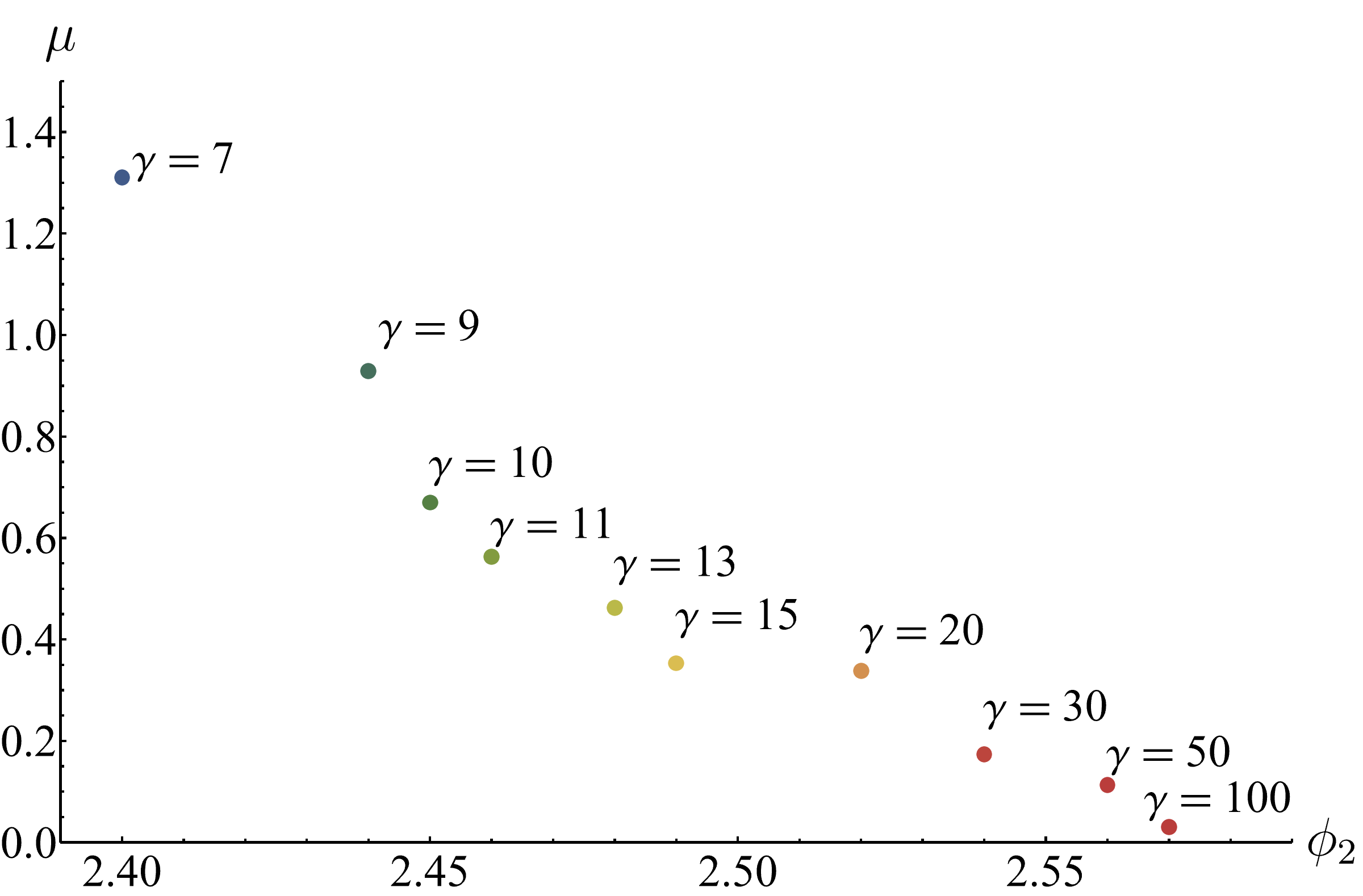}
    \captionsetup{justification=raggedright,singlelinecheck=false}
    \caption{The values of the dimensionless factor int the    surface free energy density of the bubble as the function of $\gamma$}
    \label{fig:mu}
\end{figure}

To get the canonical normalization of the kinetic term for $\varsigma$ we must multiply it on the factor,
\begin{equation}
\tilde{\Psi}_c\sim \frac{\pi}{\sqrt{2N}}z_H^2\Psi_c\sim \frac{\pi}{\sqrt{2N}}z_H^2\varsigma.
\end{equation}
Then the WKB asymptotics for $\tilde{\mu}$ is given by \cite{Hindmarsh:2020hop,Rubakov:2017xzr},
\begin{equation}
\tilde{\mu}\sim\frac{\pi}{C\sqrt{2N}}z_H^2\int\limits_0^{\varsigma_\text{min}}
            d\varsigma \,
            \sqrt{2 \left.\Big(U_\text{eff}(\varsigma)
                - U_\text{eff}(0)
            \Big)\right|_{\phi_2 = \phi_2^\text{crit}}}\nonumber
            \sim\frac{3\sqrt{3}\pi^{7/2}}{2}C\frac{T^3}{{v_4}}\mu,
\end{equation}
where we used \eqref{SigmaNorm}, \eqref{EffectivePotential} and $z_H=\frac{1}{\pi T}$. We also introduced the factor,
\begin{equation}\label{tension}
\mu=\int\limits_0^{\langle\sigma\rangle_\text{min}}
            d\sigma \,
            \sqrt{2 \mathcal{F}(\sigma)
            \Big|_{\phi_2 = \phi_2^\text{crit}}}.
\end{equation}
For this asymptotics to be valid we need ${v_4} \ll 1$, i.e. the quartic coupling should be weak. We depict the values of $\mu$ for different values of $\gamma$ at Fig. \ref{fig:mu}.

\section{Gravitational waves spectrum}
    The parameters of our model are restricted by the experimental bounds on the mass of the lowest predicted particle i.e. the radial fluctuations of the condensate
    (\ref{CondensateOperator}).
    To obtain its mass we consider the small inhomogeneous radial fluctuation
    on the homogeneous background (\ref{PTDescription:BackgroundField})
    \begin{equation}
        \chi(z) \to \chi(z) + \delta\chi(t,\vec x, z), \quad
        |\delta\chi| \ll |\chi|.
    \end{equation}
    The first-order perturbation equation on the correction takes form
    \begin{multline}
        - \frac{z^2}{f(z)} z_H^2\partial_t^2 \delta \chi
        + z^2 z_H^2 \vec\nabla^2 \delta\chi
        + z^5e^{-\phi_2z^2}\partial_{z}\Big(
            \frac{1}{z^3}e^{\phi_2z^2}f(z)\partial_{z}\delta \chi
        \Big)
        \\
        + 3\Big(1 + 3\chi^2-5\gamma\chi^4\Big) \delta \chi = 0.
    \end{multline}
  
    Being linear equation on $\delta\chi$, it can be factorised on plane waves
    $\delta\chi = e^{iEt}e^{i \vec p \vec x}u(z)$.
    For low temperatures corresponding to high values of $\phi_2$ we must consider the region near the conformal boundary $z \to 0$, where the Lorentz symmetry and the energy-momentum relation $m^2 = E^2 - p^2$ is restored. After the substitution $u(z) = e^{-\phi_2 z^2/2}z^{3/2}y(z)$ that removes the first derivative term the equation takes the Schr{\"o}dinger-like form,
    \begin{equation}
        - \frac{1}{2}y'' + \Big(\frac{3}{8z^2}+\frac{(\phi_2^2-6)z^2}{2}\Big)y = \left(\frac{m^2z_H^2}{2} + \phi_2\right) y
    \end{equation}
    where only terms up to $O(z^2)$ are taken into account. For large $\phi_2$ the potential corresponds to a harmonic oscillator with a reflecting wall at $z=0$. This gives $2m^2z_H^2\simeq \phi_2$ and, correspondingly $2m^2\simeq \tilde \phi_2$ for the lowest state, which we will use in our following considerations.

\begin{figure}[h]
        \centering
        \includegraphics[width=0.45\textwidth]{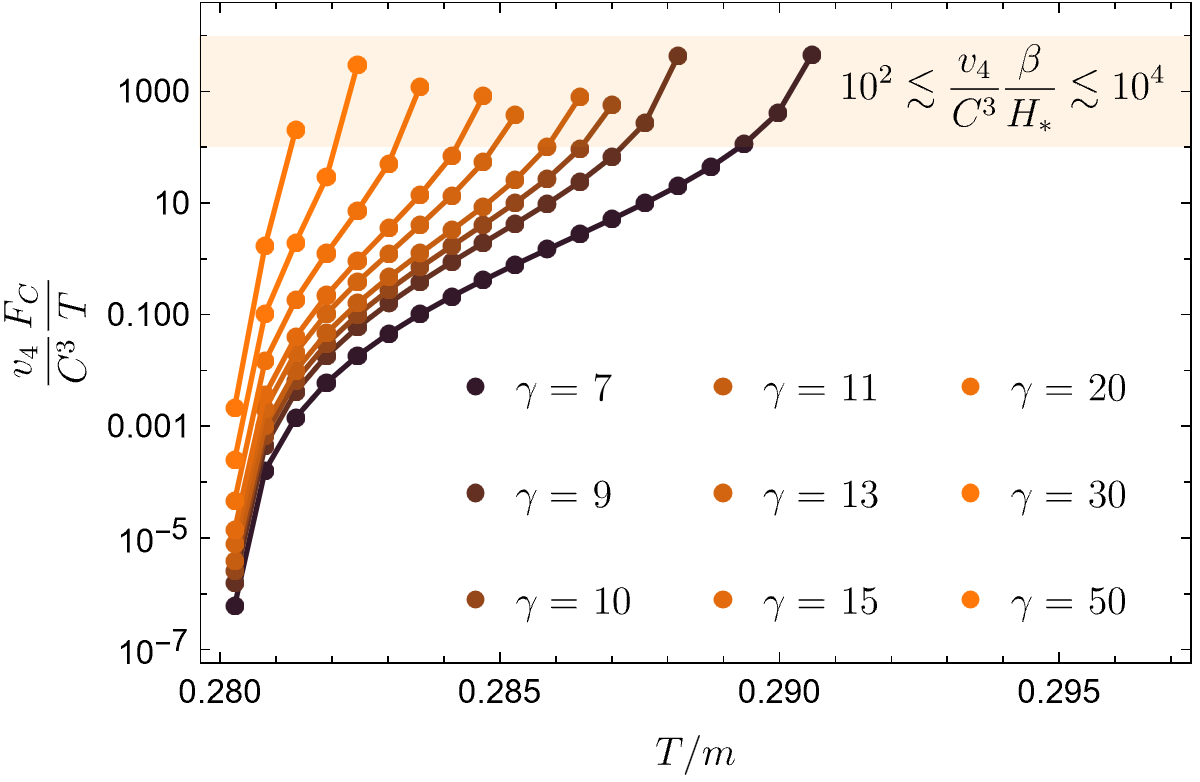}
        \captionsetup{justification=raggedright,singlelinecheck=false}
        \caption{Normalized critical value of the bubble free energy
        for different values of the parameter $\gamma$.}
        \label{fig:FCritBubble}
    \end{figure}

    The bound $m \gtrsim 10$ TeV
    ties the temperature $T = m/\pi \sqrt{2/\phi_2}$ to the physical scales and allows us to consider its cosmological implications.
    The critical value of the bubble free energy takes form
    \begin{equation}
        F_C = F_\text{bubble}(R_\text{(crit)})
        = \frac{3}{2}\sqrt{3\pi^3}\frac{C^3}{{v_4}}\frac{\mu^3}{\mathcal{F}^2}T
    \end{equation}
    with $F_\text{in} = 0$ and $dF_\text{out}/dV = U_\text{eff}$.
    Its numerical values are presented in the picture Fig.\ref{fig:FCritBubble}

    The bubble collision during the nucleation phase results in the gravitational wave production.
    The time-depended nucleation rate is given by the formula
    \cite{Kamionkowski:1993fg}
    \begin{equation}
        \Gamma(t) = \Gamma_* e^{\beta (t - t_*)},
    \end{equation}
    where $1/\beta$ is a constant timescale from nucleation to initial collision.
    The nucleation takes place at time $t = t_*$ when the nucleation rate becomes comparable with the Hubble rate
    $H_*^4 \Gamma \sim 1$.
    At the radiation dominant stage the Hubble rate is determined by the number of the relativistic degrees of freedom $g_*$ (which is close to the number of the SM degrees of freedom $\sim 100$) by $H_* = \sqrt{90/(8\pi^3 g_*)} M_\text{Pl}$.
    
    We can neglect the impact of the cosmic expansion on the nucleation if $\beta/H_* \gg 1$.  
    This ratio can be estimated from the bubble free energy as follows,
    \begin{equation}
        \frac{\beta}{H_*} = T \partial_T \log \Gamma
        \sim \frac{F_C}{T}.
    \end{equation}
    The bubble production starts at the critical temperature
    $T_\text{crit.} = m/\pi \sqrt{2/\phi_2^\text{crit.}}$
    (see Fig. \ref{fig:phasediag}) corresponding to the upper points of the graphs
    in the picture Fig. \ref{fig:FCritBubble}.
    The values of $(v_4/C^3) \beta/H_*$
    do not depend on $\gamma$
    and justify our approximation of slow universe expansion. For $v_4 \lesssim 10^{-1} \ll 1$ and $C \sim 1$ we get $10^3 \lesssim \beta/H_* \lesssim 10^5$.

    The parameter $\alpha = \rho_0 / \rho_\text{rad.}$ 
    \cite{Breitbach:2018kma} determines the liberated energy during the phase transition.
    The corresponding energy densities are~\footnote{
    $\rho_0$ can be found as the deviation of the effective potential
    (\ref{EffectivePotential}) from critical point
    $U_\text{eff} - U_\text{eff}^\text{crit.}$
    within analytical solutions (\ref{Analytical:SigmaAndDilaton})
    and (\ref{Analytical:FreeEnergyDensity}).
    }
    \begin{equation}
        \rho_\text{rad.} \sim g_* T^4, \quad
        \rho_0 = \frac{6\pi}{v_4} \pi^2 T^2 m^2 \frac{\Delta T}{T},
    \end{equation}
    where $\Delta T=T_{(0)}-T_\text{crit.}$.
    As one can see from the Fig. \ref{fig:FCritBubble},
    $\Delta T/T$ is the range
    $10^{-2} \lesssim \Delta T/T(\gamma) \lesssim 10^{-1}$
    for the considered range of $\gamma$.
    Therefore, the released energy is of order
    \begin{equation}
        \alpha \sim \frac{20}{v_4} \frac{\Delta T}{T}
        = \frac{2}{v_4} (10^{-1}\text{--}10^0) \gtrsim 2.
    \end{equation}

    The main contribution in the run-away spectrum of the 
    gravitational waves gives the scalar part produced
    during initial collisions of the bubble walls
    \cite{Breitbach:2018ddu}.
    Sound and turbulence contributions are not currently
    included in the rough estimate.
    The peak of the spectrum can be found as
    \cite{Jinno:2017fby}
    \begin{equation}
        \Omega_\text{GW} h^2 =
        1.67 \cdot 10^{-5} \kappa \Delta
        \left(\frac{\beta}{H_*}\right)^{-2}
        \left(\frac{\alpha}{1 + \alpha}\right)^2
        \left(\frac{g_*}{100}\right)^{-\frac{1}{3}}.
    \end{equation}
    Where the efficiency factor $\kappa(\alpha) \sim 1$ for $\alpha\gg 1$, the velocity factor $\Delta$ may be approximated by $\Delta = 0.2 k/\beta$ with the wall momentum $k \lesssim \beta$
    (in the relativistic approximation)\cite{Jinno:2017fby}.

    The corresponding peak frequency can be found as follows \cite{Jinno:2017fby},
    \begin{equation}
        f_0 = 1.65 \cdot 10^{-5} \cdot \frac{1}{2\pi}\frac{k}{\beta} 
        \frac{\beta}{H_*}\frac{T}{0.1 \text{TeV}}
        \left(\frac{g_*}{100}\right)^\frac{1}{6} \text{ Hz}.
    \end{equation}

\begin{figure}[h]
    \centering
    \includegraphics[width=\textwidth]{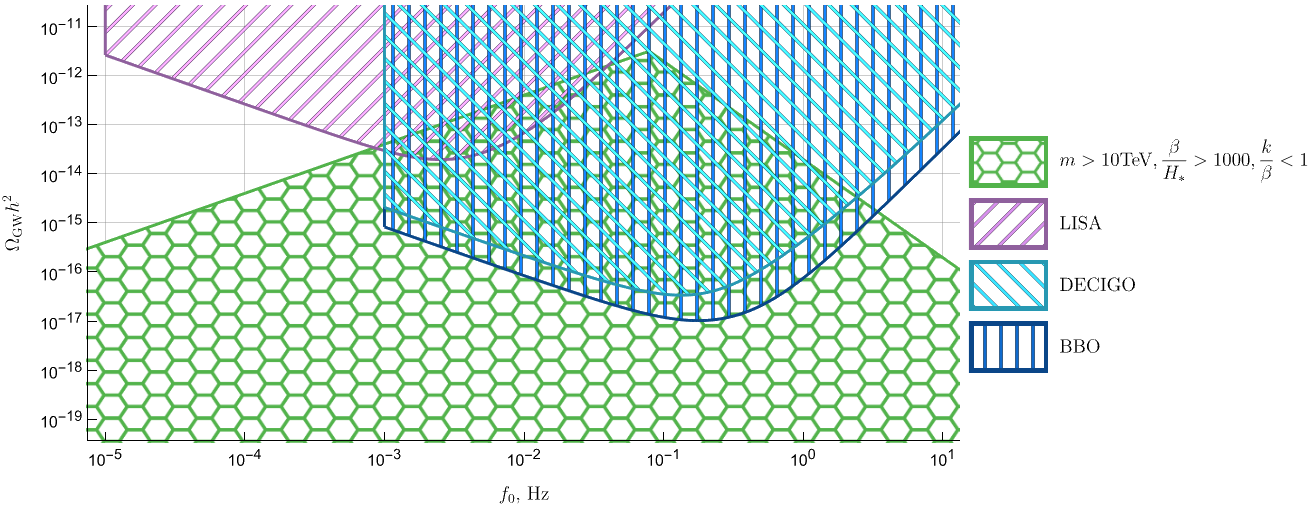}
    \captionsetup{justification=raggedright,singlelinecheck=false}
    \caption{The green region is the predicted (peak of the spectrum) values
    of the gravitational waves at the corresponding frequency.
    Sensitivity curves of the perspective gravitational waves observatories
    (LISA, DECIGO and BBO) are also placed on the figure \cite{Schmitz:2020syl}.}
    \label{fig:GWSpectrum}
\end{figure}

    The estimated gravitational wave background is contrasted with the capabilities of the future detectors on Fig. \ref{fig:GWSpectrum}.
    The right high-frequency boundary for our predicted region (green field)
    is due to the restriction $k < \beta$.
    The left low-frequency boundary is determined by the experimental restriction on the lowest composite state masses $m \gtrsim 10$ TeV and by $\beta/H_* \gtrsim 1000$.

    Most of existing gravitational observatories are sensitive only to higher frequencies and higher amplitudes. However, our estimates lie within the sensitivity range of some of the next generation gravitational observatories. Note that Ultimate DECIGO is not shown as it covers most of the shown region with the sensitivity
    $\Omega_\text{GW} h^2 \sim 10^{-19}$ \cite{Braglia:2021fxn}.

\section{Conclusions}

In this work we have constructed the dynamical holographic model that experiences the first order phase transition from the symmetric phase at high temperatures to the broken symmetry phase at low temperatures and have estimated the rate of the bubble nucleation. We employed the perturbation theory to study the phase diagram in the semi-analytic fashion. This approach helps to gain more intuitive understanding of the relation between the shape of the five-dimensional potential and the resulting phase diagram.
Besides, we estimate the gravitational waves spectrum
produced due to first order phase transition.
Our estimates imply that such background should be detectable with the planned gravitational waves observatories.

These results may have importance not only for the physics beyond the Standard model but also for the holographic description of the QCD at high densities where the first order phase transition should occur \cite{Guenther:2020jwe}. The rich physics may be associated with the emergence of the bubbles of the hadronic and quark matter, e.g. local $\mathcal{CP}$ violation \cite{Kharzeev:1998kz,buckley2000can,kharzeev2006parity,Andrianov:2009pm,Andrianov:2007kz,Kovalenko:2020ryt,Vioque-Rodriguez:2021hgp,Novikov:2022pff}.

Our paper leaves a number of important questions for the further investigation. The shape of the potential should be matched with the properties of the corresponding correlation functions in the dual gauge theory, and we, basically, pointed out what relation should four-point and six-point corellators satisfy for the first order phase transition to occur. We also have left the possibility of more complex interactions such as the amplitude proportional to the 't Hooft determinant like the one considered in \cite{fang2019chiral}. The Einstein-dilaton sector that is responsible for the confinement-deconfinement transitions should also be taken into account. Last but not the least, is the interaction with the weakly coupled sector of the Standard model fields that determines how the processes considered in this work are related to the observable quantities.

\section*{Acknowledgements}
We would like to thank Sergey Afonin and Alexander Andrianov and Dmitry Ageev 
for the discussions and suggestions. This research was funded by the Russian Science Foundation grant number
\makebox{21-12-00020}.

\bibliographystyle{unsrt}
\bibliography{holhiggs_phase}

\end{document}